# Co-existence of Micro, Pico and Atto Cells in Optical Wireless Communication

Osama Zwaid Alsulami
*School of Electronic and Electrical Engineering*
*University of Leeds*
Leeds, United Kingdom
ml15ozma@leeds.ac.uk

Mohamed O. I. Musa
*School of Electronic and Electrical Engineering*
*University of Leeds*
Leeds, United Kingdom
m.musa@leeds.ac.uk

Mohammed T. Alresheedi
*Department of Electrical Engineering*
*King Saud University*
Riyadh, Saudi Arabia
malresheedi@ksu.edu.sa

Jaafar M. H. Elmirghani
*School of Electronic and Electrical Engineering*
*University of Leeds*
Leeds, United Kingdom
j.m.h.elmirghani@leeds.ac.uk

*Abstract*— Interference between cells or users can have a significant impact on the quality of optical wireless communication (OWC) links. This paper studies the co-existence of infrared based Micro cells with Visible light communication based Pico and Atto Cells for downlink communication. The signal to noise ratio (SNR) of each cell and the signal to interference and noise ratio (SINR) between cells are evaluated when an angle diversity receiver and an imaging receiver are used. The results show that Atto cell systems provide higher SNR and data rates compared to the Pico cell systems which provide higher SNR and data rates compared to Micro cell systems. The Micro cell systems however provide mobility as they provide a larger coverage area, followed by the Pico cell systems.

## I. INTRODUCTION

With the dramatic increase in the number of users using communication networks, there is a growing demand for increasing the bandwidth in wireless communication systems. In addition, the growth in high data rates leads researchers to seek another form of wireless communication systems that can meet these demands. Currently the radio frequency (RF) spectrum is the preferred technology for wireless communication in an indoor environment. However, because of the limited available radio spectrum, the channel capacity and transmission rate will be limited and this can affect the performance of the communication systems. Also, providing very high data rates over 10 Gbps and up to the Tbps region by using the available radio spectrum bandwidth is challenging. Cisco believes that the Internet traffic between 2016 and 2021 is going to increase 27 times [1]. Under this expectation, the quest for alternative spectrum bands that can support these demands beyond radio spectrum is underway. Optical wireless communication (OWC) is a promising communication technique for exchanging data at high data rates. Data rates up to 20 Gbps in an indoor environment can be achieved for transferring videos, data and voice by using OWC systems [2]-[11]. In addition, license free bandwidth, low-cost, and high security can be provided by using OWC systems compared to RF wireless systems [12]-[19]. However, some limitations can appear such as the absence of the line-of-sight (LOS) link and the resulting considerable decrease in the system's performance [20], [21]. Also, inter-symbol interference (ISI) which is caused by multi-path propagation can affect the system. Interference from different sources in OWC systems can affect the system's performance as well.

In this paper, the co-existence of the Micro, Pico and Atto Cells in OWC is studied. An Infrared and 'red, yellow, green, and blue (RYGB) Laser Diodes (LDs)' are used as OWC transmitters. In addition, an angle diversity receiver is used and the developed model considers the effects of mobility and multi-path propagation. A white colour can be provided by using RYGB LDs. This can be used for indoor illumination as stated in [22], and for transmission to provide a high modulation bandwidth. This paper is organised as follows: The system configuration is described in Section II and the optical receiver design is shown in Section III. Section IV introduces the simulation results, and the conclusion are provided in Section V.

## II. SYSTEM CONFIGURATION

Three Optical wireless cellular systems are assumed to co-exist in a room that has a (length × width × height) dimensions equal to 8 m × 4 m × 3 m as shown in Figure 1. On the ceiling of the room, 3 meters high, the different transmitters are located. The first cellular system uses a single infrared transmitter at the centre of the ceiling with a wide semi angle equal to 65° to cover the whole room, representing a Micro cell (see Figure 2a). While, the second and third systems use 8 angle diversity transmitter (ADT) light units, as Pico and Atto cells, placed uniformly across the roof at (1 m, 1 m, 3 m), (1 m, 3 m, 3 m), (1 m, 5 m, 3 m), (1 m, 7 m, 3 m), (3 m, 1 m, 3 m), (3 m, 3 m, 3 m), (3 m, 5 m, 3 m) and (3 m, 7 m, 3 m) as shown in Figure 1. Each ADT light unit uses 5 branches and each branch is oriented to different location based on the Azimuth (Az) and Elevation (El) angles [8]. The *El* angle of the branch that faces downward is set at -90° while the other branches are set to -70°. The *Az* angle of the branch that faces downwards is set to 0° and the other branches are chosen to be 45, 135, and 225. The Pico cell system uses the branch that faces downward with semi angle equal to 40° to divide the communication floor into 8 cells with dimensions (2 m × 2 m) (see Figure 2b). Whereas, the Atto cell system uses the other four branches that have semi angle equal to 21° to cover a small cell in the communication floor with dimensions (1 m × 1 m) (see Figure 2c).

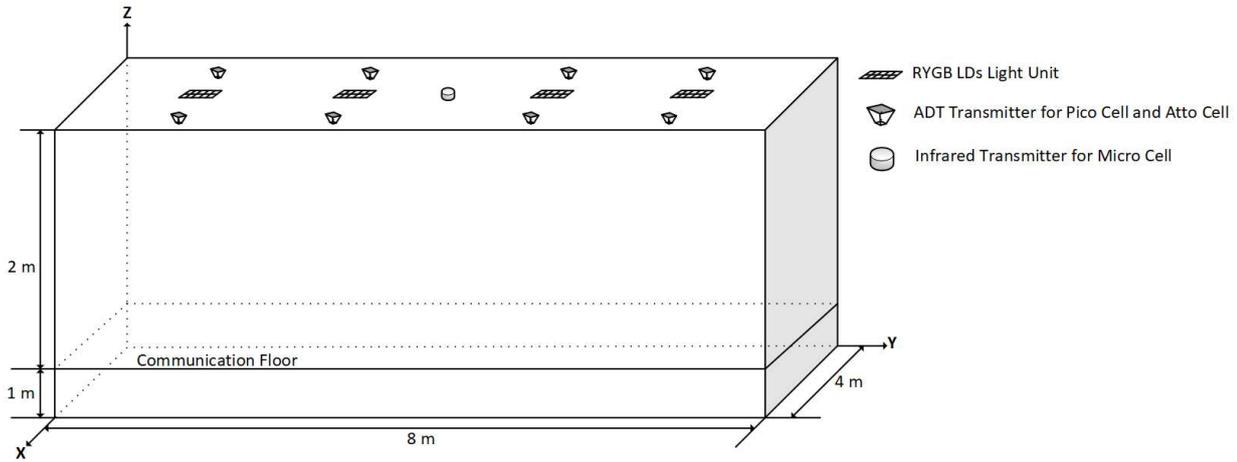

Fig. 1. Room Configuration

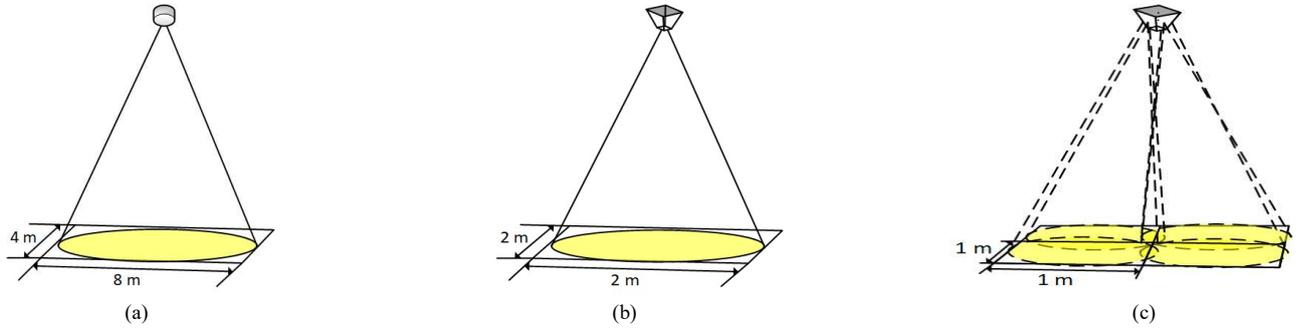

Fig. 2. Micro cell configuration, (b) Pico cell configuration, (c) Atto cell configuration.

An empty room that has neither doors nor windows is used in the simulation. In this work, up to second order reflections were considered. The third and higher order reflections have little impact on the received power [23], [24]. A ray-tracing algorithm is utilised to model reflections from the ceiling, walls and the floor in the room. Thus, each surface in the room is divided into small equal areas with size $dA$, and a reflection coefficient of $\rho$. A plaster wall reflects light rays close to a Lambertian pattern as shown in [24]. Thus, each surface in the room ceiling, walls and floor is modelled as a Lambertian reflector with coefficient of reflection equal to 0.8 for ceiling and walls and 0.3 for the floor. Each element in each surface acts as a small emitter that reflects the received ray in the form of the Lambertian pattern with $n$ (emission order of the Lambertain pattren) equals to 1. The area of the surface elements can play a significant role in the resolution of the results. When the surface elements are very small, higher temporal resolution results are produced. However, this can come at the cost of long computation time of the simulation. Therefore, 5 cm × 5 cm was chosen as the area of the surface element for the first order reflection, while 20 cm × 20 cm was chosen as the area of the surface element for the second order reflection to keeping the computation time of the simulation within a reasonable time [18], [25]. The communication floor (CF) is set at 1 m above the floor as shown in Figure 1 which means all communications are done above the CF.

When using the proposed ADT light units, the required illumination (lx) cannot be achieved based on the ISO and European illumination requirements [26]. Thus, four extra light units have been used for illumination. Each light unit consist of 10 wide semi angles RYGB LDs with beams set at 70° to provide an acceptable light level inside the room. These light units are placed at different location in the ceiling (2 m, 1 m, 3 m), (2 m, 3 m, 3 m), (2 m, 5 m, 3 m) and (2 m, 7 m, 3 m). A minimum illumination level was achieved equal to 306.4 lx which is just above the minimum requirement of 300 lx. In addition, the maximum level of illumination is not exceeded (1300 lx) as shown in Figure 3.

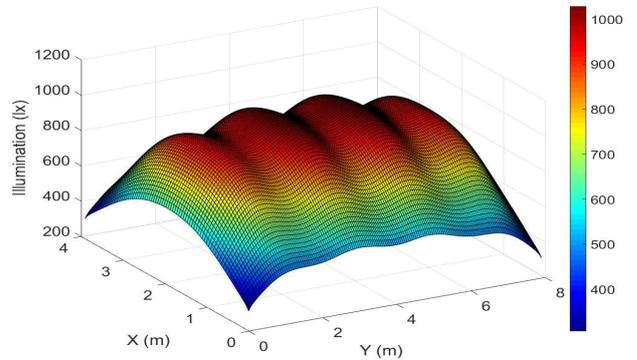

Fig. 3. The illumination (lx) inside the room

## III. Optical Receiver Design

An angle diversity receiver (ADR) with 7 branches of narrow FOV photodetectors (See Figure 4) is used to reduce interferences. Each photodetector is oriented at a different direction to cover a different transmitter in the ceiling based on the elevation and azimuth angles. The $El$ angles of the seven detectors set as follow: six detectors are set at 40°, while the detector that faces upwards is positioned at 90°. Whereas, the $Az$ angles of detectors are 0°, 60°, 120°, 180°, 240°, 300°, and 0°, respectively [27]. The FOV of the six detectors is set to 25°, while the detector that faces upwards has a 30° FOV. In addition, the area of each photodetector is $4\ mm^2$ with responsivity of 0.4 A/W.

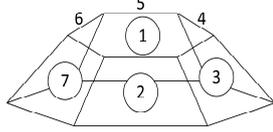

Fig. 4. Angle Diversity Receiver (ADR) configuration.

## IV. Simulation Setup and Results

To meet the bit error rate requirements, the ADR receiver is designed to produce a signal to noise ratio that corresponds to the required BER using equation 1 for the OOK system, given as:

$$BER = Q(\sqrt{SINR}), \qquad (1)$$

where $Q(\cdot)$ is the Q-function which is expressed as follows:

$$Q(x) = \frac{1}{2} erfc\left(x/\sqrt{2}\right) \approx \frac{1}{\sqrt{2\pi}} \frac{e^{-\left(x^2/\sqrt{2}\right)}}{x} \qquad (2)$$

The Signal to interference and noise ratio for the three coexisting systems is given as:

$$SINR = \frac{R^2\ (P_{s1}-P_{s0})^2}{\sigma_t^2 + \sum_{c=1}^{C} R^2 (P_{i1}-P_{i0})^2} \qquad (3)$$

where $c$ is the set of systems used and can take two values referring to the other two cell systems interfering with the current operating cell system, $R$ is the photodetector responsivity ($R = 0.4\ A/W$), $P_{s1}$ is the received signal power associated with logic 1, $P_{s0}$ is the received signal power associated with logic 0, $P_{i1}$ is the power received from the other interfering cells associated with logic 1, $P_{i0}$ is the power received from the other interfering cells associated with logic 0 and $\sigma_t$ is the total noise associated with the received signal and is computed by:

$$\sigma_t = \sqrt{\sigma_{pr}^2 + \sigma_{bn}^2 + \sigma_{sig}^2} \qquad (4)$$

where, $\sigma_{pr}$ is related to the preamplifier noise, $\sigma_{bn}$ is related to the background noise and $\sigma_{sig}$ is the noise related to the received signal. Selection combining (SC) and Maximum ratio combining are both evaluated and compared. For the selection combining approach, the SINR is given as:

$$SINR_{SC} = max_k \left(\frac{R^2\ (P_{s1}-P_{s0})^2}{\sigma_t^2 + \sum_{c=1}^{C} R^2 (P_{i1}-P_{i0})^2}\right)_k, \qquad 1 \le k \le J \qquad (5)$$

where $J$ is the total number of detectors.

The SINR for the maximum ratio combining (MRC) technique which combines all the outputs from all the detectors is given by:

$$SINR_{MRC} = \sum_{k=1}^{J} \left(\frac{R^2\ (P_{s1}-P_{s0})^2}{\sigma_t^2 + \sum_{c=1}^{C} R^2 (P_{i1}-P_{i0})^2}\right)_k, \qquad 1 \le k \le J \qquad (6)$$

Different receivers have been used in this work based on the supported channel bandwidth of each system. In the Micro cell system, 30 MHz receiver bandwidth has been used [28], while 1 GHz receiver bandwidth was used in the Pico cell system, and 5 GHz receiver bandwidth was utilised in the Atto cell system [29].

Figure 5a shows the SNR result of the three systems when they are used independently of each other. For these levels of SNR, the Micro cell system achieves a data rate up to 42.8 Mbps. While, the Pico cell system provides a data rate up to 1.4 Gbps, and the Atto cell system supports a data rates up to 7.1 Gbps. The smaller distance between the receiver and the transmitter can provide a high received power which can increase the SNR as shown in Pico and Atto cells. However, the Micro cell system can support room wide mobility, compared to the other systems. Figure 5b shows the gain in SNR (dB) when the MRC is used for each of the three cells independently. Micro cell can achieve 3 dB gain at the sides of the room and this gain diminishes closer to the centre of the room where the transmitter is located. However, for the Atto and Pico cells, the gain reaches around 6 dB but is at a low value of 4 dB at locations that are closest to the transmitters.

Figure 5c compares the SINR between the three cells systems, at different combinations. The results show that when there is interference between cells, the SINR is reduced especially between small cells due to the high received power from the interfering cell. Figure 5d illustrates the gain in SINR (dB) in the three systems of cells when MRC is utilised. Using Atto cells with Micro cells and Pico cells with Micro cells, a 6 dB gain can be achieved. However, this is reduced by around 4 dB when the receiver is placed between the Atto and Pico cells transmitters' locations. Whilst, when applying all systems of cells together or small cells system together, the gain reaches around 4.5 dB between small cells transmitters' locations.

However, MRC shows a better result compared to the other techniques due to its ability to combine optimally all the detectors' outputs as shown in Figure 5b and 5d.

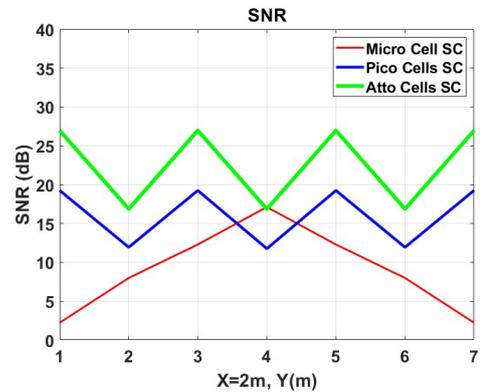

(a)

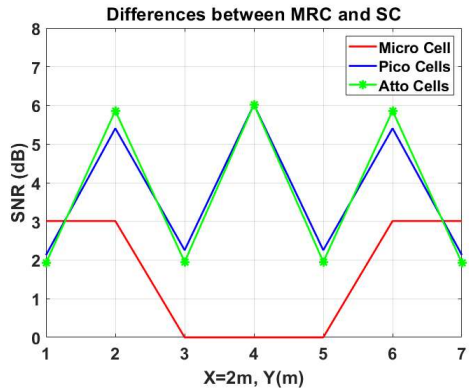

(b)

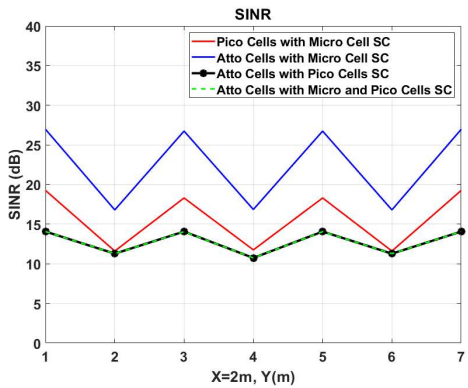

(c)

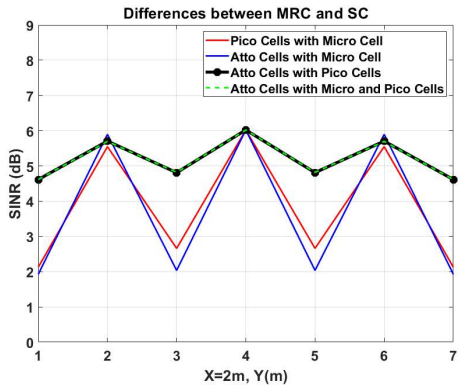

(d)

Figure 5: (a) SNR of each cell system, (b) Differences between MRC and SC in SNR, (c) SINR between cells, (d) Differences between MRC and SC in SINR.

## V. CONCLUSIONS

Three communications systems configurations based on OWC were proposed in this paper: Micro, Pico and Atto cell OWC systems. The Micro cell system consists of one transmitter that covers the communication floor. While, the Pico and Atto cell systems include 8 ADT LDs units. The Pico cell system uses the branch that faces downward to cover 2 m × 2 m of the communication floor. Whereas, the Atto cell system uses the other four branches and each branch covers a small area of the communication floor that is 1 m × 1 m. The proposed systems were examined by using an angle diversity receiver (ADR). When the cell becomes small, the achieved data rate and SINR increase as the Atto cell system shows in this work compared to other systems. Moreover, the interference between the Atto cell and Micro cell systems is very low which can provide a high SINR at high data rate. However, the Micro cell system can support mobility compared to the other systems.

ACKNOWLEDGMENT

The authors would like to acknowledge funding from the Engineering and Physical Sciences Research Council (EPSRC) for the TOWS project (EP/S016570/1). The first author would like to thank Umm Al-Qura University in the Kingdom of Saudi Arabia for funding his PhD scholarship. All data are provided in full in the results section of this paper.